\documentclass[twocolumn,showpacs,preprintnumbers,amsmath,amssymb]{revtex4}
\usepackage{graphicx}
\usepackage{dcolumn}
\usepackage{bm}

\begin{document}

\title{Influence of absorbing dielectric background on bistable response of dense collection of two-level atoms}

\author{Denis V. Novitsky}
\email{dvnovitsky@tut.by}
\affiliation{%
B.I. Stepanov Institute of Physics, National Academy of Sciences of
Belarus, \\ Nezavisimosti~Avenue~68, 220072 Minsk, Belarus.
}%

\date{\today}

\begin{abstract}
The stationary problem of light interaction with a dense collection
of two-level atoms embedded in a dielectric is treated
semiclassically. The effect of dielectric absorption on the
phenomenon of intrinsic optical bistability is discussed in terms of
the complex local-field enhancement factor. The calculations
accounting propagation effects (nonlocal regime) show that even low
dielectric absorption results in elimination of response
instabilities.
\end{abstract}


\maketitle

\section{Introduction}

The effect of optical bistability is the corner-stone of optical
information processing. It is the basis for implementation of such
devices as all-optical switches, transistors, and logical gates. The
straightforward method to obtain the bistable response is to place a
nonlinear medium inside a feedback system, for example, Fabry--Perot
resonator \cite{Gibbs}. In the late 1970s Bowden and Sung introduced
\cite{Bowd79} the idea of the mirrorless (intrinsic) optical
bistability as a result of interatomic (dipole-dipole) correlations.
This idea was then developed in a number of papers (see, for
example, \cite{Hopf, BenAryeh, Friedberg}).

The usual starting point in the analysis of intrinsic optical
bistability (IOB) is the semiclassical generalized Bloch equations
\cite{Bowd93} which take into account the near-dipole-dipole
interactions of two-level atoms (the so-called local field
correction). However, there is a problem of influence of the
background (host) dielectric medium on the bistable response.
Crenshaw proved \cite{Cren08} that the semiclassical approach gives
the correct explanation of local-field enhancement by the linear
dielectric in accordance with the \textit{microscopic} quantum
electrodynamics (QED). Interestingly, this enhancement cannot be
accounted for by the \textit{macroscopic} QED. Therefore, if the
process of spontaneous emission is not to be examined, we can use
the semiclassical Bloch equations.

The local-field enhancement due to the host dielectric is measured
by the factor $\ell=(\varepsilon+2)/3$, where $\varepsilon$ is the
dielectric permittivity. Since $\varepsilon$ can be complex, we can
automatically take into account dielectric absorption. It was shown
in Ref. \cite{Cren96} that the real part of $\ell$ reduce the IOB
threshold while the imaginary part acts in the opposite direction.
In one of our previous publications \cite{Nov08} we studied the
effects of local and nonlocal bistability and internal coherent
reflection under the assumption of absence of the background
dielectric influence. In this short paper we try to fill this
omission.

The paper is divided into several parts. Section II is devoted to
the solution of the stationary problem of monochromatic radiation
interaction with the dense collection of two-level atoms in a
dielectric. We analyze the influence of dielectric properties on the
main parameters of local bistability. In Section III we consider the
overall permittivity of such a media and its bistable response.
Finally, in Section IV the problem of radiation propagation in a
layer of finite thickness is discussed. The effect of dielectric on
the so-called nonlocal bistability is numerically analyzed.

\section{Bistable response}

\begin{figure}[t!]
\includegraphics[scale=0.9, clip=]{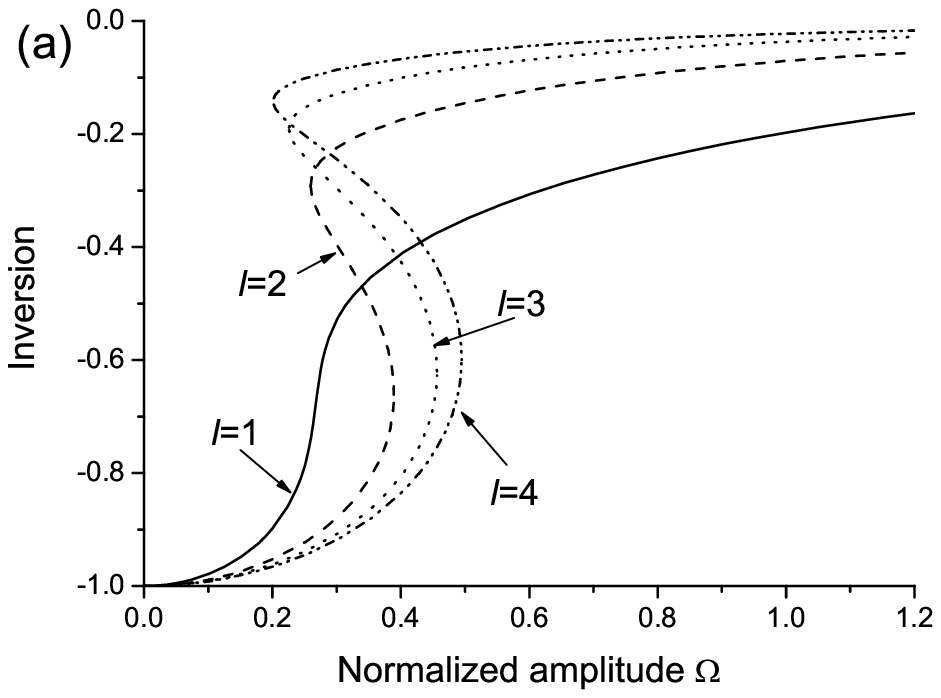}
\includegraphics[scale=0.9, clip=]{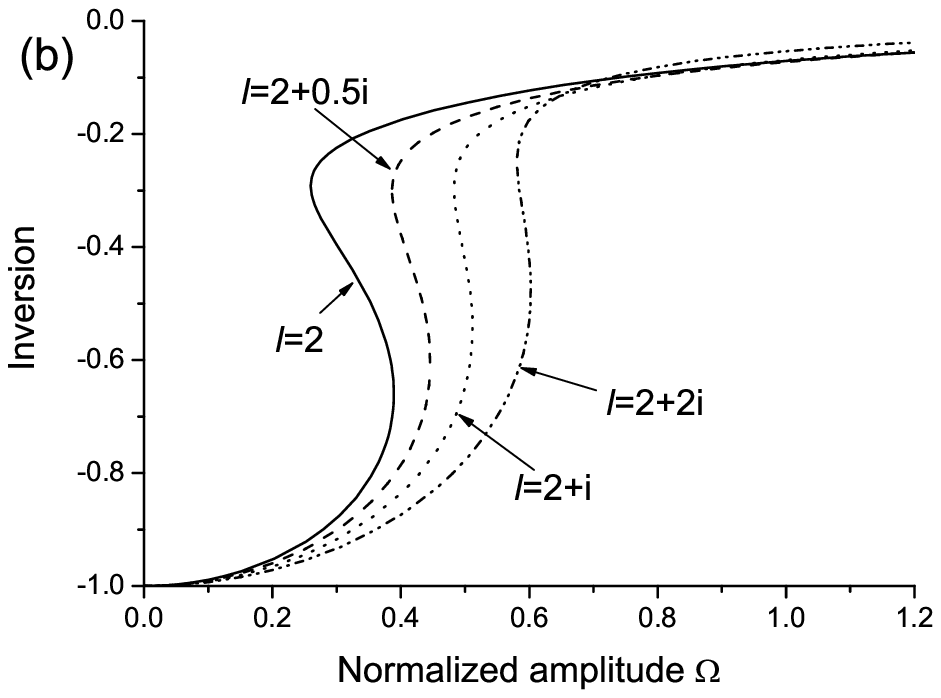}
\caption{\label{fig1} Dependence of inversion on the light amplitude
at (a) real values of factor $\ell$, (b) complex values of factor
$\ell$. Calculation parameters: $W_{\textrm{eq}}=-1$, $\delta=-2$,
$b=4$, $\Gamma=0.1$.}
\end{figure}

\begin{figure}[t!]
\includegraphics[scale=0.9, clip=]{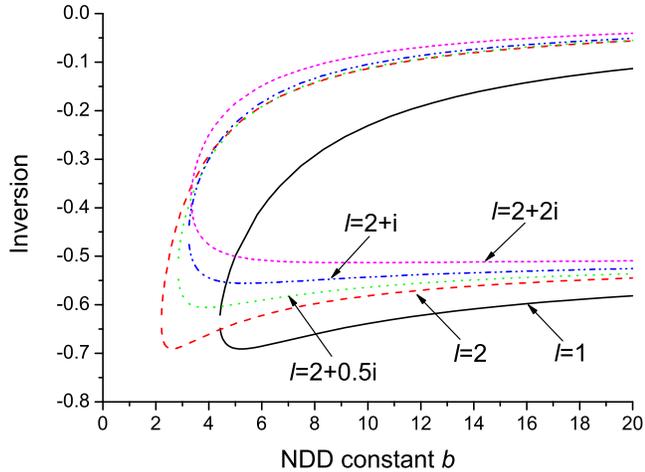}
\caption{\label{fig2} (Color online) Positions of the inflection
points as a function of the NDD parameter $b$. Other parameters:
$W_{\textrm{eq}}=-1$, $\delta=-2$.}
\end{figure}

\begin{figure}[t!]
\includegraphics[scale=0.9, clip=]{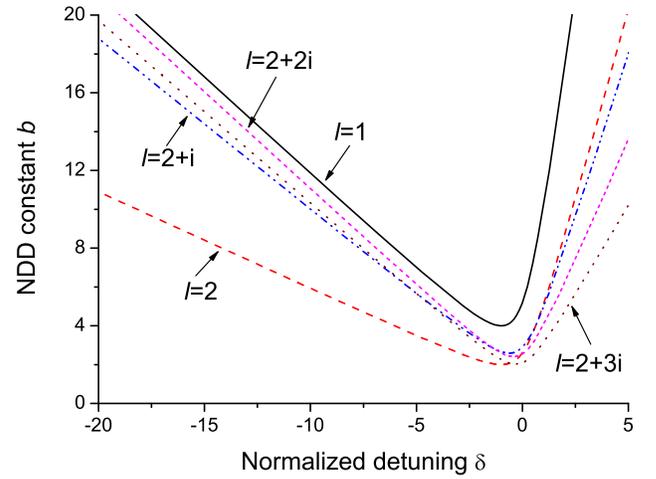}
\caption{\label{fig3} (Color online) The region of bistability
existence in the plane ($\delta$, $b$).}
\end{figure}

\begin{figure}[t!]
\includegraphics[scale=0.9, clip=]{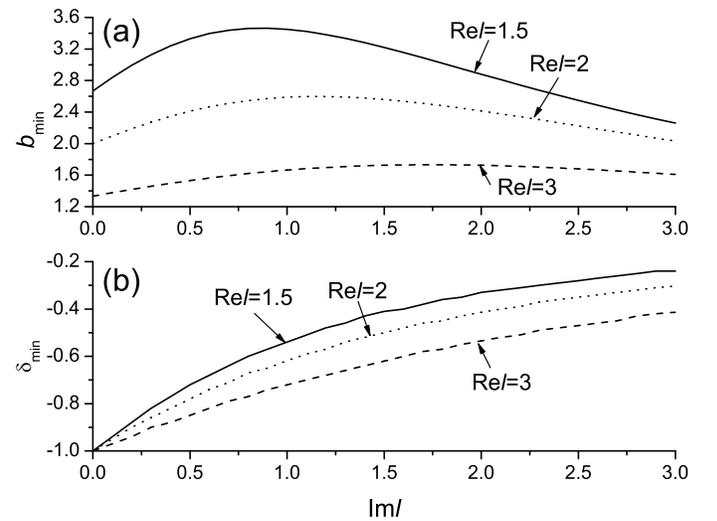}
\caption{\label{fig4} The dependence of minimal NDD parameter value
(a) and corresponding detuning (b) on the imaginary part of the
local-field enhancement factor $\ell$.}
\end{figure}

In the paper we consider the normal incidence of linearly polarized
monochromatic plane wave on a dense resonant medium. We start with
semiclassical Bloch equations which take into account near
dipole-dipole (NDD) interactions of two-level atoms as well as
background dielectric effect \cite{Cren96, Cren08},
\begin{eqnarray}
\frac{\partial R}{\partial t}&=& -i \frac{\mu}{2 \hbar} \ell E W + i (\Delta-\ell \epsilon W) R - \gamma_2 R, \label{dPdt} \\
\frac{\partial W}{\partial t}&=& -i \frac{\mu}{\hbar} (\ell^*
E^* R - \ell E R^*) - 2 i (\ell^* - \ell) \epsilon |R|^2 \nonumber \\
&-& \gamma_1 (W-W_{\textrm{eq}}), \label{dNdt}
\end{eqnarray}
where $W$ is the inversion (population difference of two levels),
$R$ is the atomic polarization, $E$ is the amplitude of macroscopic
electric field, $\mu$ is the transition dipole moment, $\Delta$ is
the detuning of radiation from resonance, $\gamma_1$ and $\gamma_2$
are the population and polarization relaxation rates respectively,
$W_{\textrm{eq}}$ is the value of inversion at equilibrium. The
parameter
\begin{eqnarray}
\epsilon=\frac{4 \pi N \mu^2}{3 \hbar} \label{eps}
\end{eqnarray}
is responsible for NDD interaction strength, $N$ is the density of
two-level atoms per unit volume, $\hbar$ is the Planck constant.
Here $\ell=(\varepsilon+2)/3$ is the local-field enhancement factor
due to polarizability of host material with dielectric constant
$\varepsilon$ (generally, complex). Appearance of factor $\ell$
results in three local-field effects \cite{Cren96}: (i) enhancement
of the magnitude and the phase shift (if $\varepsilon$ is complex)
of the electric field; (ii) Lorentz frequency shift
($\omega_L=\epsilon W$) enhancement due to real part of $\ell$;
(iii) cooperative decay due to imaginary part of $\ell$ in both
equations (\ref{dPdt}) and (\ref{dNdt}).

In stationary regime we obtain the expression for the polarization,
which can be written in dimensionless notation as follows
\begin{eqnarray}
R=\frac{1}{2} \frac{i \ell W \Omega}{i (\delta-\ell b W)-1},
\label{polar}
\end{eqnarray}
where $\Omega=\mu E/\hbar \gamma_2$, $\delta=\Delta/\gamma_2$,
$b=\epsilon/\gamma_2$. It is seen that the polarization depends on
the inversion which can be calculated from the equation
\begin{eqnarray}
\Gamma (W-W_{\textrm{eq}}) \left| i (\delta-\ell b W)-1 \right|^2+W
|\ell \Omega|^2=0. \label{eqinv}
\end{eqnarray}
Here $\Gamma=\gamma_1/\gamma_2$.

The cubic equation (\ref{eqinv}) describes the phenomenon of
intrinsic optical bistability (IOB) well known from the previous
researches. However, the presence of dielectric background changes
the quantitative characteristics and conditions of IOB. As can be
seen from Fig. \ref{fig1}(a), the Lorentz shift enhancement (the
second effect named above) results in easier observation of bistable
response. In vacuum case ($\ell=1$), if the other parameters are the
same, there is no bistability at all. On the other hand, if the
dielectric is absorptive [complex factor $\ell$, see Fig.
\ref{fig1}(b)], bistable loop not only shifts towards higher light
intensities, but also gets more and more narrow.

This implies that absorption of the background material changes the
region of IOB existence somehow. Let us consider this question in
detail. The necessary condition for the Eq. (\ref{eqinv}) to
demonstrate bistability is that the equation $d|\Omega|^2/dW=0$ has
two different roots in the physically appropriate range $-1 \leq W
\leq 0$. This equation can be written in explicit form as
\begin{eqnarray}
2 b^2 |\ell|^2 W^3 &-& b \left(2 \textrm{Im} \ell + 2 \delta
\textrm{Re}
\ell + b |\ell|^2 W_{\textrm{eq}} \right) W^2 \nonumber \\
&+& W_{\textrm{eq}} (\delta^2+1)=0. \label{eqderiv}
\end{eqnarray}
Two solutions of Eq. (\ref{eqderiv}) in the range $-1 \leq W \leq 0$
represent the inflection points of bistability which are seen in the
Fig. \ref{fig1} (the points of jump from one solution to another).
When these points coincide, the bistability loop disappears. Figure
\ref{fig2} shows the behavior of the inflection points as NDD
parameter is decreasing. The critical (minimal) value of $b$ at
which bistability still exists decreases for larger real parts of
the local-field enhancement factor and increases for larger
imaginary parts of $\ell$. At the same time the inversion
corresponding to the jumps between the solutions is growing.

This means that the region of parameters $\delta$ and $b$, where the
bistability, occurs changes as well. Some examples of these regions
of bistability existence are demonstrated in Fig. \ref{fig3}: above
the curves plotted the bistability is present. In the case of vacuum
($\ell=1$) we obtain the curve that was reported previously
\cite{Nov08}. The minimal value of the NDD parameter in this case is
$b_{\textrm{min}}=4$ for the detuning $\delta_{\textrm{min}}=-1$. If
$\ell$ increases and is still real, the region of bistability is
getting wider. For example, for $\ell=2$ we have
$b_{\textrm{min}}=2$ at the same value of the detuning
$\delta_{\textrm{min}}=-1$. However, if the background dielectric is
absorptive ($\ell$ is complex), this minimal value of NDD parameter
increases.

From Fig. \ref{fig3} one can see that $b_{\textrm{min}}$ for
$\textrm{Im}\ell=1$ is larger than for $\textrm{Im}\ell=0$. However,
for $\textrm{Im}\ell=2$ it is already smaller than in the former
case. Detailed study of the behavior of the minimal value of the NDD
parameter [Fig. \ref{fig4}(a)] shows that there is a maximum in
dependence $b_{\textrm{min}} (\textrm{Im}\ell)$. As
$\textrm{Im}\ell$ tends to infinity, the $b_{\textrm{min}}$ is
slowly decreasing down to zero. At the same time the value of
detuning, corresponding to the minimal NDD parameter, monotonically
tends to zero, too. Note that, for smaller $\textrm{Re}\ell$, the
curve $b_{\textrm{min}} (\textrm{Im}\ell)$ has more pronounced
maximum and then slower converge to zero. In contrast to this,
$\delta_{\textrm{min}} \rightarrow 0$ faster for smaller
$\textrm{Re}\ell$.

Moreover, as $\textrm{Im}\ell$ is growing, the bistability existence
region, at first, gets much more narrow for $\delta <
\delta_{\textrm{min}}$ and slightly wider for $\delta >
\delta_{\textrm{min}}$ (see Fig. \ref{fig3}). But then it tends to
get wider in both cases. In the limit $\textrm{Im}\ell \rightarrow
\infty$, the region of bistability occupies all the half-plane
$b>0$.

\section{Dielectric permittivity}

\begin{figure}[t!]
\includegraphics[scale=0.9, clip=]{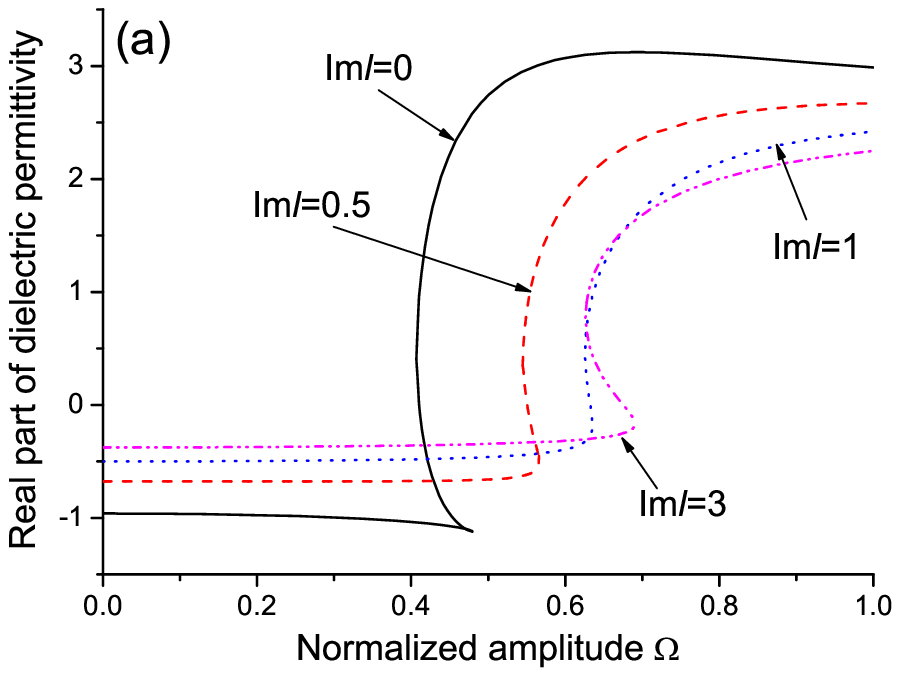}
\includegraphics[scale=0.9, clip=]{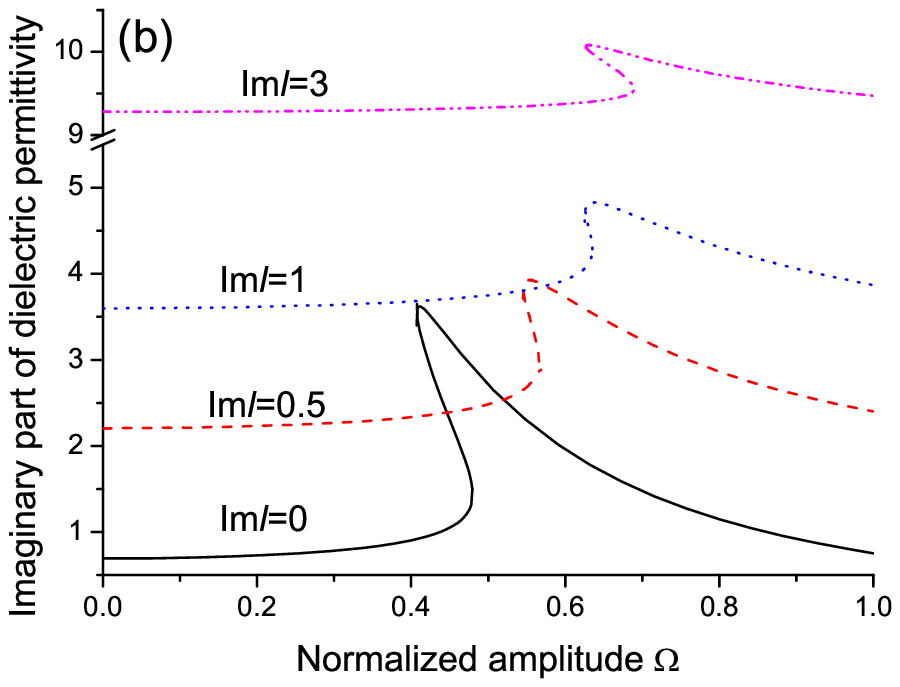}
\caption{\label{fig5} Dependence of (a) real and (b) imaginary parts
of the dielectric permittivity on the light amplitude. Calculation
parameters: $W_{\textrm{eq}}=-1$, $\delta=-1$, $b=4$, $\Gamma=0.1$,
$\textrm{Re} \ell=1.5$, $\textrm{Im} \ell$ is a varying parameter.}
\end{figure}
\begin{figure}[t!]
\includegraphics[scale=0.9, clip=]{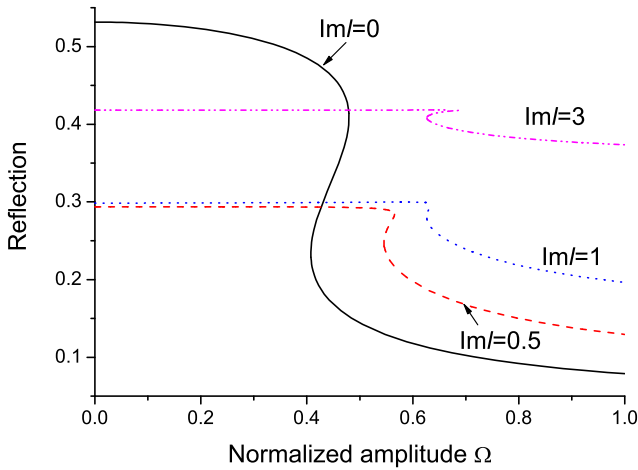}
\caption{\label{fig6} Dependence of reflection at the vacuum-medium
interface on the light amplitude. Calculation parameters are the
same as in Fig. \ref{fig5}.}
\end{figure}

The resulting dielectric permittivity of the dielectric doped with
two-level atoms can be written as
\begin{eqnarray}
\varepsilon_{\textrm{total}}=1+4 \pi \chi, \label{epstot}
\end{eqnarray}
where $\chi=P/E$ is the susceptibility, $P$ is the total
polarization defined as a sum of the linear polarization of the
background dielectric (permittivity $\varepsilon$) and the nonlinear
polarization $P_{\textrm{res}}=2 N \mu R$ due to resonant atoms, so
that \cite{Cren96}
\begin{eqnarray}
P=\frac{\varepsilon-1}{4 \pi} E + \frac{\varepsilon+2}{3}
P_{\textrm{res}}. \label{polartot}
\end{eqnarray}
Substituting the stationary expression (\ref{polar}) to the
equations (\ref{polartot}) and (\ref{epstot}) we obtain
\begin{eqnarray}
\varepsilon_{\textrm{total}}=\varepsilon + \frac{3 i \ell^2 b W}{i
(\delta - \ell b W) - 1}. \label{epstot1}
\end{eqnarray}
It is worth to recall that $\ell=(\varepsilon+2)/3$.

Expression (\ref{epstot1}) with cubic equation (\ref{eqinv}) defines
the bistable behavior of the dielectric permittivity (Fig.
\ref{fig5}). This bistability, in accordance with the results of the
previous section, is getting more and more narrow at first and then
wider, as the imaginary part of local-field enhancement factor is
growing. Moreover, it is seen that the real part of
$\varepsilon_{\textrm{total}}$ appears to be negative at
low-intensive branch of the bistability. This negativity corresponds
to the well-studied effect of the internal coherent reflection
\cite{Mal95, Mal97, Nov08}. This phenomenon can be obtained in pure
form if the imaginary part of $\varepsilon_{\textrm{total}}$ is
absent. If it were so, then the condition
$\textrm{Re}\varepsilon_{\textrm{total}}<0$ would lead to the unit
value of the reflection coefficient, i.e.
$|(\sqrt{\varepsilon_{\textrm{total}}}-1)/(\sqrt{\varepsilon_{\textrm{total}}}+1)|^2=1$.
However, as Fig. \ref{fig5}(b) shows,
$\textrm{Im}\varepsilon_{\textrm{total}}$ increases for greater
$\textrm{Im} \ell$. As a result, the reflection for large
$\textrm{Im}\varepsilon_{\textrm{total}}$ is far from unity and it
only slightly differs for both branches of the bistability (see Fig.
\ref{fig6}). The difference between the branches tends to decrease
for $\textrm{Re}\varepsilon_{\textrm{total}}$ and
$\textrm{Im}\varepsilon_{\textrm{total}}$ as well.

\section{Nonlocal bistability (Propagation effects)}

\begin{figure}[t!]
\includegraphics[scale=0.9, clip=]{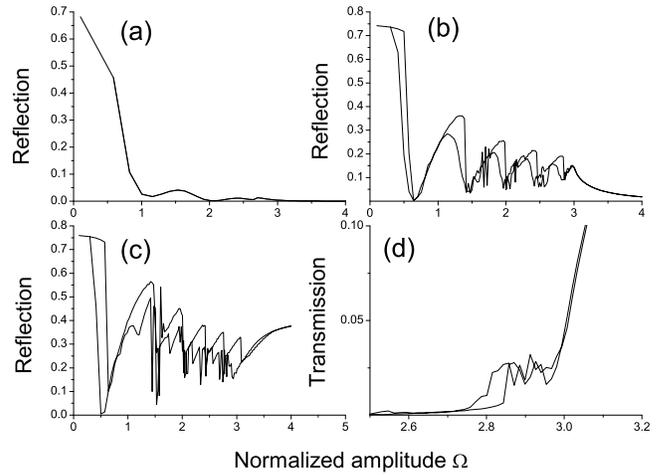}
\caption{\label{fig7} Dependence of reflection (a, b, c) and
transmission (d) on the light amplitude. Calculation parameters:
$W_{\textrm{eq}}=-1$, $\delta=-1$, $b=4$, $\Gamma=0.1$, light
wavelength $\lambda=0.5$ $\mu$m. The enhancement factor (a)
$\ell=1$, (b, d) $\ell=2$, (c) $\ell=3$. The thickness of the layer
is $L=0.5$ $\mu$m.}
\end{figure}

\begin{figure}[t!]
\includegraphics[scale=0.9, clip=]{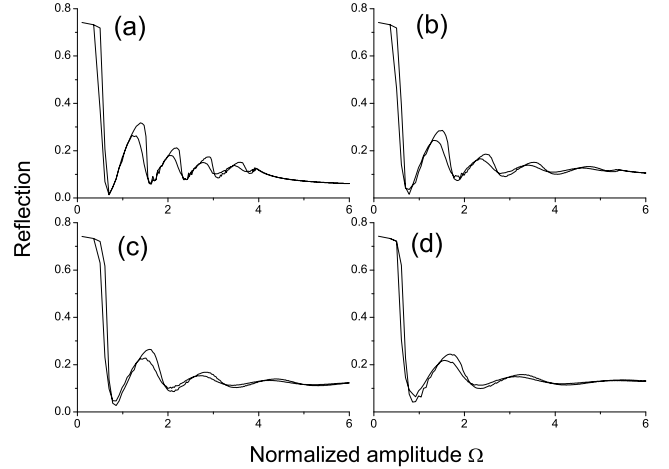}
\caption{\label{fig8} Dependence of reflection on the light
amplitude. The enhancement factor (a) $\ell=2 + 0.1 i$, (b) $\ell=2
+ 0.2 i$, (c) $\ell=2 + 0.3 i$, (d) $\ell=2 + 0.4 i$. The other
parameters are the same as in Fig. \ref{fig7}.}
\end{figure}

Now let us consider the layer of the two-level medium with finite
thickness. In this case we have to take into account the light
propagation effects resulting in the so-called nonlocal optical
bistability \cite{Mal97}. To analyze propagation of monochromatic
light wave in the nonlinear layer, we use the iteration matrix
method which was described in detail in Ref. \cite{Nov08}. This
method is the modification of the widely known transfer matrix
approach. Due to this method, we divide the layer into sublayers
with thicknesses much less than the light wavelength so that the
properties of each sublayer can be treated as constant. Then,
according to initial conditions (refractive index of unexcited
medium), we calculate the distribution of light inside the layer
and, hence, obtain new values of refractive index of each sublayer.
These calculations can be continued many times until, at a certain
iteration, the distribution of light and refractive index take on
the stationary form (with a certain accuracy).

The results of calculations of reflection and transmission
coefficients are demonstrated in Fig. \ref{fig7} and \ref{fig8}. In
Fig. \ref{fig7} one can see one more confirmation of the role of the
local-field enhancement factor $\ell$ in bistability appearance.
There is also multiplicity of hysteresis loops that can be treated
as the main feature of nonlocal bistability \cite{Nov08}. The number
of loops depends on the layer thickness and the value of $\ell$
[compare figures \ref{fig7}(b) and (c)]. Transmission has only one
loop corresponding to the most intensive reflection loop [Fig.
\ref{fig7}(d)].

Even small absorption of host dielectric significantly changes this
bistable response as shown in Fig. \ref{fig8}. As the imaginary part
of $\ell$ grows, the loops tend to stretch along the light amplitude
axis. The number of loops and their width are decreasing as well.
There is another important consequence of background absorption
which is connected with the stability of the effect. As one can see
in Fig. \ref{fig7}(b) and (c), reflection at intensities between the
loops demonstrate many spikes distributed chaotically. These spikes
correspond to the regions of misconvergence of the iteration matrix
method so that we have to stop it at a certain iteration and take
next intensity value. As opposed to this behavior, the regions of
bistability have rather fast convergence. Perhaps, the spikes are
due to instability by auto-oscillations mechanism reported in
\cite{Mal95, Mal97}. Figure \ref{fig8} shows that even low level of
absorption leads to elimination of spikes, i.e. response of the
system becomes stable. This means that the two-level medium with the
background dielectric of quite low absorption can be used to obtain
only slightly suppressed multiple bistability without
auto-oscillations.

\section{Conclusion}

In this article we have considered semiclassically the general
situation of the dense collection of two-level atoms placed into
absorbing background dielectric. Of course, the coefficient of
dielectric absorption cannot take on arbitrary values. We also do
not take into account the problem of medium heating due to radiation
absorption. Nevertheless, the analysis carried out above can be
useful for proper selection of the media in possible experiments. In
particular, the appropriate choice of slightly absorbing (and,
hence, slightly heated) background medium can help to obtain the
hysteresis response without such instabilities as auto-oscillations.


\begin{thebibliography}{25}
\bibitem{Gibbs} H. M. Gibbs, \textit{Optical bistability: Controlling light with light} (Academic Press Inc., 1985).
\bibitem{Bowd79} C. M. Bowden and C. C. Sung, \pra {\bf19}, 2392 (1979).
\bibitem{Hopf} F. A. Hopf, C. M. Bowden, and W. H. Louisell, \pra {\bf29}, 2591 (1984).
\bibitem{BenAryeh} Y. Ben-Aryeh, C. M. Bowden, and J. C. Englund, \pra {\bf34}, 3917 (1986).
\bibitem{Friedberg} R. Friedberg, S. R. Hartmann, and J. T. Manassah, \pra {\bf40}, 2446 (1989).
\bibitem{Bowd93} C.M. Bowden and J.P. Dowling, \pra {\bf47}, 1247 (1993).
\bibitem{Cren08} M. E. Crenshaw, \pra {\bf78}, 053827 (2008).
\bibitem{Cren96} M. E. Crenshaw, \pra {\bf53}, 1139 (1996).
\bibitem{Nov08} D. V. Novitsky and S. Yu. Mikhnevich, \josab {\bf25}, 1362 (2008).
\bibitem{Mal95} V. Malyshev and E. C. Jarque, \josab {\bf12}, 1868 (1995).
\bibitem{Mal97} V. Malyshev and E. C. Jarque, \josab {\bf14}, 1167 (1997).
\end{thebibliography}
\end{document}